\documentclass[11pt,tightenlines,superscriptaddress]{revtex4-2}
\usepackage{amsmath,amssymb}
\usepackage{mathptmx}
\usepackage{hyperref}
\usepackage{url}
\usepackage{graphicx}
\usepackage{makecell}

\newcommand{\vect}[2][]{\ensuremath{\mathbf{#2}_\text{#1}}}
\newcommand{\uvec}[2][]{\ensuremath{\mathbf{\hat{#2}}_\text{#1}}}

\newcommand{\rmsoffs}{\ensuremath{\left\lVert\theta_\text{offs}\right\rVert_2}}

\begin{document}

\title{Design of an arrangement of cubic magnets for a quasi-axisymmetric 
       stellarator experiment}


%
\author{K.~C.~Hammond}
\email{khammond@pppl.gov}
\affiliation{Princeton Plasma Physics Laboratory, Princeton, NJ 08543, USA}
\author{C.~Zhu}
\affiliation{Princeton Plasma Physics Laboratory, Princeton, NJ 08543, USA}
\affiliation{Present affiliation: Department of Plasma Physics and Fusion
             Engineering, School of Nuclear Science and Technology, 
             University of Science and Technology of China, 
             Hefei 230026, China}
\author{K.~Corrigan}
\affiliation{Princeton Plasma Physics Laboratory, Princeton, NJ 08543, USA}
\author{D.~A.~Gates}
\affiliation{Princeton Plasma Physics Laboratory, Princeton, NJ 08543, USA}
\author{R.~Lown}
\affiliation{SABR Enterprises, LLC, North Andover, MA 01845, USA}
\author{R.~Mercurio}
\affiliation{SABR Enterprises, LLC, North Andover, MA 01845, USA}
\author{T.~M.~Qian}
\affiliation{Princeton Plasma Physics Laboratory, Princeton, NJ 08543, USA}
\author{M.~C.~Zarnstorff}
\affiliation{Princeton Plasma Physics Laboratory, Princeton, NJ 08543, USA}



\begin{abstract}
The usage of permanent magnets to shape the confining magnetic field of 
a stellarator has the potential to reduce or eliminate the need for 
non-planar coils. As a proof-of-concept for this idea, we have developed a
procedure for designing an array of cubic permanent magnets that works in 
tandem with a set of toroidal-field coils to confine a stellarator plasma.
All of the magnets in the design are constrained to
have identical geometry and one of three polarization types in order to simplify
fabrication while still producing sufficient field accuracy. We present
some of the key steps leading to the design, including the geometric 
arrangement of the magnets around the device, the procedure for
optimizing the polarizations according to the three allowable magnet types,
and the choice of magnet types to be used. We apply these methods to design
an array of rare-Earth permanent magnets that can be paired with a set of 
planar toroidal-field coils to confine a quasi-axisymmetric plasma with a 
toroidal magnetic field strength of about 0.5~T on axis.
\end{abstract}

\maketitle

%
%
%
%
%

\section{Introduction}
\label{sec:intro}

The stellarator is a magnetic fusion reactor concept that seeks to confine a 
burning plasma with a precisely-shaped, non-axisymmetric magnetic field. 
The most advanced stellarator designs to date have typically relied on
complex, non-planar coils to generate this field. Since non-planar coils
have been a major driver of cost and construction time for stellarator
experiments \cite{nielson2010a,rummel2012a,bosch2013a}, methods for achieving 
desired magnetic fields with simpler coil geometry have been a major area of 
research \cite{pomphrey2001a,brown2015a,gates2017a,zhu2018a,landreman2017a,
paul2018a,lobsien2018a,kruger2021a,lonigro2022a,giuliani2022a}.
Recently, permanent magnets were proposed as an approach to coil 
simplification \cite{helander2020a}.
In principle, permanent magnets can contribute to the three-dimensional
shaping of the magnetic field, thereby easing the shaping requirements for
any non-planar coils or even eliminating the need for non-planar coils
altogether.

Many strategies for designing suitable magnet arrays
have been developed. Some approaches entailed calculating a spatial distribution
of magnetization throughout a three-dimensional layer surrounding a winding
surface that encloses the target plasma \cite{zhu2020a,landreman2021a,xu2021a}.
These procedures typically employ a Fourier representation of the 
magnetization distribution and as such share much in common with methods for
designing modular coils for stellarators \cite{merkel1987a,landreman2017a}.
Other approaches optimize the dipole moments of discrete magnets in arbitrary
locations \cite{zhu2020b,lu2021a,lu2022a,qian2022a}. 

In all of these cases, the distribution of magnetization $\mathbf{M}$ through 
the magnet arrangement is determined through an optimization procedure.
The spatial distribution of $\mathbf{M}$ is defined parametrically, e.g.
through Fourier harmonics or through the moments of individual dipoles in the
arrangement. The optimization procedure adjusts these parameters to
minimize the normal component of the total magnetic field on the boundary
of the targeted plasma equilibrium, including contributions from the permanent
magnets, toroidal-field (TF) coils, any additional coils, and plasma 
currents.

To reduce the complexity and cost of magnet fabrication, it is preferable to
minimize the number of unique magnet types required for an arrangement, both
in terms of geometry and magnetization. Formulating and imposing such 
constraints on the optimization procedure is a nontrivial endeavor, and one 
that admits multiple possible strategies. One approach has been to constrain
the polarization of each magnet to be locally normal to a winding surface that 
encloses the plasma and conforms to the plasma's boundary 
geometry \cite{zhu2020a,xu2021a,lu2021a}. In such a scheme, the magnets
could be cut from thin slabs of uniformly-polarized magnetic material and
oriented to be locally normal to the winding surface. The drawback of such
a constraint is that the solution tends to make less efficient use of the
magnets than solutions that find the optimal polarization direction at each
location. Hence, while this approach works well for target plasmas with
relatively low three-dimensional field shaping requirements
\cite{zhu2020a,lu2021a,qian2022a}, a greater degree of freedom in the
polarization direction may be necessary if stronger shaping fields are 
required or if engineering constraints limit the amount of magnet material 
that can be placed near the plasma boundary \cite{hammond2020a}.

For magnet array designs in which the polarization direction is optimized,
the magnet shapes and their spatial layout can be more easily decoupled from 
a toroidal winding surface. This enables the use of simpler and fewer different 
magnet shapes and mounting structures. However, to limit the total number
of unique magnet types required, the polarization cannot be allowed to
vary continuously for each magnet and must instead be chosen from a discrete
set of allowable polarizations. If each magnet is constrained to have a 
cubic shape with a polarization along an axis of symmetry perpendicular to
two of its faces, this would enable solutions in which only one unique magnet 
type is required, as recently demonstrated by Lu et al.~\cite{lu2022a}.
An alternative method for optimizing magnet arrays with discrete polarizations
will be presented in this paper.

In this work, we utilize these advances in discrete magnet optimization to 
design a magnet array for a mid-scale quasi-axisymmetric stellarator in which
the magnetic field is created with permanent magnets and planar TF coils. The
array consists of three unique types of magnets, each with identical cubic
geometry and distinguished only by its polarization orientation.
The magnets are designed to confine a variant of the
``C09R00'' equilibrium, originally designed for the NCSX 
experiment \cite{zarnstorff2001a,nelson2003a}, with the magnetic field on axis 
scaled down to about 0.5~T. Correspondingly, the magnet array will be 
constrained to fit around the NCSX vacuum vessel. In addition, it will be 
assumed that the NCSX TF coils are to provide the toroidal component of the
magnetic field; no other coils will be incorporated into the design.
The target plasma has a major radius of 1.44~m, a minor radius of 0.32~m,
and a volume-averaged beta of 4.1\%.

\section{Outline of the design and optimization procedure}
\label{sec:procedure}

The procedure developed in this work for specifying a magnet arrangement can be
summarized as follows:

\begin{enumerate}
    \item Generate an arrangement of cubic magnets
    \item Perform a continuous optimization of the dipole moments of each manget
          in two steps:
    \begin{enumerate}
        \item Minimize an objective function for magnetic field inaccuracy 
        \item Minimize a weighted sum of objective functions for
              field inaccuracy and intermediate dipole moment magnitudes
    \end{enumerate}
    \item Rotate each optimized moment vector to the nearest allowable 
          discrete vector
\end{enumerate}

\noindent The remainder of this paper will describe each of these steps in
more detail and present the results of the application of this procedure to
the design of a magnet array for the target plasma. 
Sec.~\ref{sec:layout} will describe the geometric arrangement of the magnets
around the plasma vessel. Sec.~\ref{sec:optimization} will describe the 
procedures used to perform continuous optimizations of the dipole moments of 
each magnet.
Sec.~\ref{sec:mag_mapping} will describe the polarization types
that were considered for the discrete solution and explain the down-selection
to the three types used for the magnet array design.

\section{Magnet layout and mounting concept}
\label{sec:layout}

The scheme for the spatial layout of the magnets seeks to pack magnets as 
tightly as possible around the plasma-conforming vacuum vessel, as magnets
positioned close to the plasma provide the required
three-dimensional magnetic field shaping more efficiently. At the same time, 
the layout must be compatible
with reasonably simple concepts for fabrication, assembly, and mounting.
As such, the layout scheme was developed in close coordination with
the design of the mounting structures.

\begin{figure*}
    \begin{center}
    \includegraphics[width=0.9\textwidth]{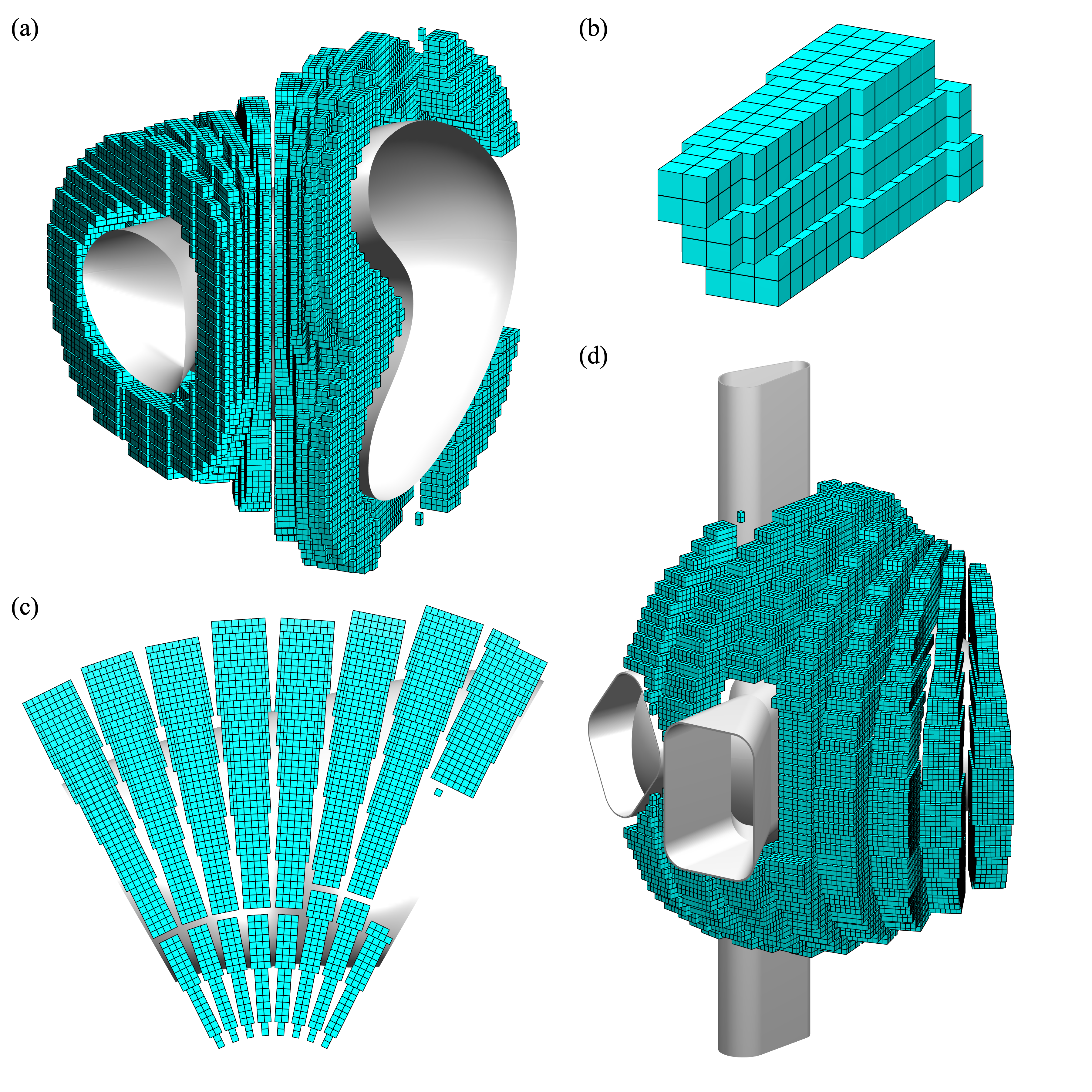}
    \caption{Arrangement of cubic magnets designed to conform to one half-period
             of the NCSX vacuum vessel, which extends $60^\circ$ toroidally. 
             (a) view from the inboard side; 
             (b) close-up view of three vertically-stacked ``drawers'' of 
                 magnets from the inboard side;
             (c) view from the top;
             (d) view from the outboard side, including ports that the magnets
                 were designed to avoid}
    \label{fig:mag_argmt}
    \end{center}
\end{figure*}

An example arrangement of cubic magnets is shown in Fig.~\ref{fig:mag_argmt}.
The magnets are divided into a set of discrete wedges, each of which subtends a 
particular toroidal angle (Fig.~\ref{fig:mag_argmt}c). 
The magnets in each wedge are partitioned into an
inboard and outboard subset, with the inboard/outboard boundaries determined
separately on the top and bottom according to proximity to the highest and
lowest points of the cross-section of the plasma vessel. Within each
partition, the magnets are subdivided into ``drawers,'' each of which is 
two layers high (Fig.~\ref{fig:mag_argmt}b). Magnets within each drawer are 
held together with glue.
The drawers are inserted in gridded structures on both the inboard and 
outboard sides of the vessel. 

The magnets were restricted to be cubic in shape in order to take advantage of
the cube's many degrees of rotational symmetry. This allows for many different 
dipole moment directions to be realized
at each location in the magnet arrangement using various orientations of a
small number of unique magnet polarization types. The edge length of the cube 
was set at 3~cm, and was limited by the fabrication procedure. Specifically,
the magnets are foreseen to be cut from a 5~cm slab of material
with a polarization orientation perpendicular to the slab plane. Hence, to
obtain a cube with a polarization that is not perpendicular to a face, it 
must be cut at an oblique angle from the slab. Our design 
utilizes three different polarization types, two of which are not 
perpendicular to a face. 3~cm is the largest possible edge length for cubes of 
each of these orientations to be cut from the slab. Our choice of these 
polarization types will be discussed in more detail in 
Sec.~\ref{sec:mag_mapping}.

\begin{figure}
    \begin{center}
    \includegraphics[width=0.4\textwidth]{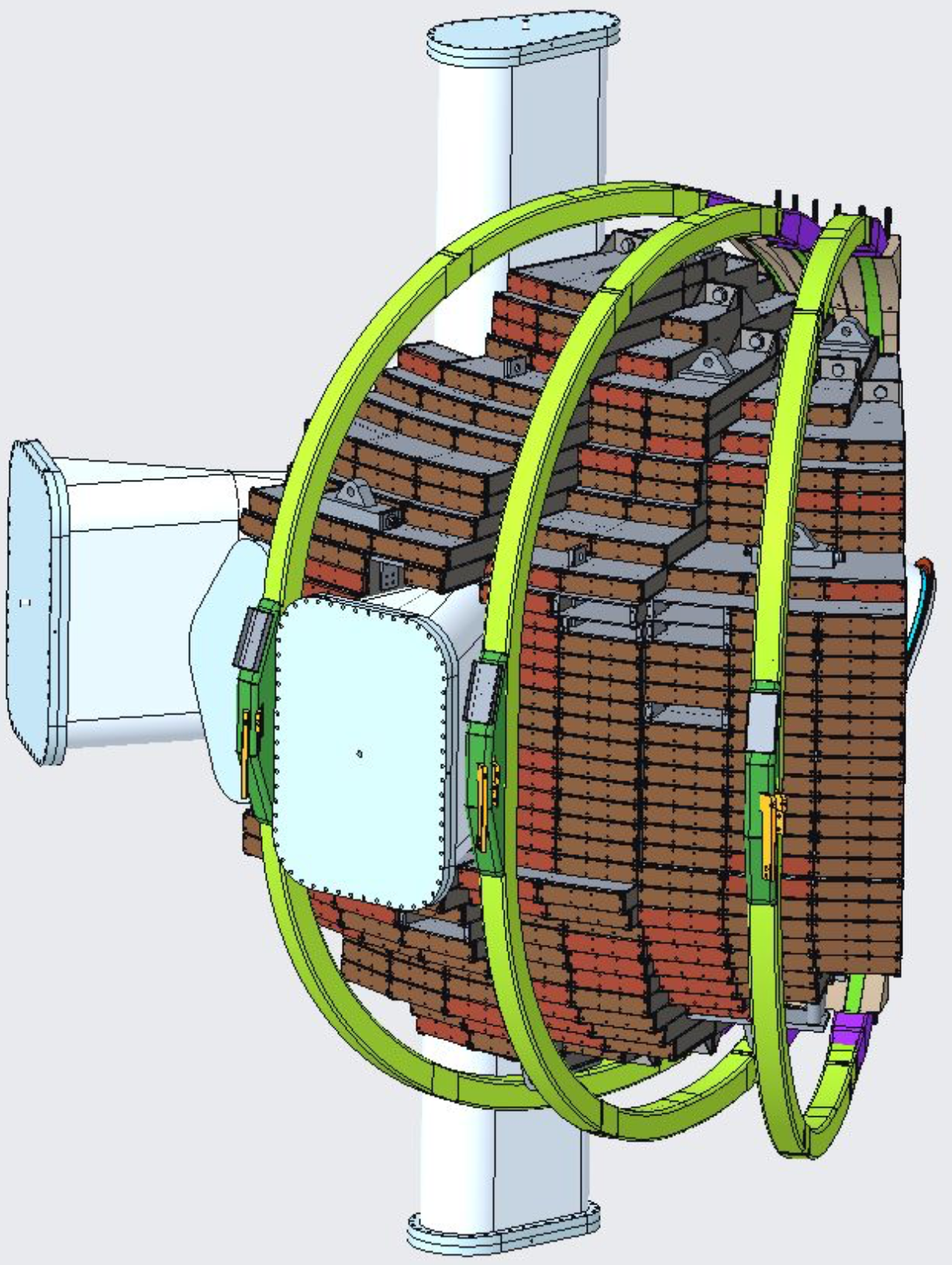}
    \caption{Rendering of the permanent magnet mounting structure positioned
             around the plasma vessel. The magnets are fully enclosed within
             this structure. The vantage point 
             for this image is similar to that of Fig.~\ref{fig:mag_argmt}d. 
             The structures fit within the bore of the toroidal-field coils,
             which are shown in yellow.}
    \label{fig:cad_structure}
    \end{center}
\end{figure}

The spacing between adjacent magnets was set to accommodate the methods for
mounting and assembly.  Magnets within the same drawer are separated by a 
distance of 0.04~cm to leave space for glue as well as a thin coating layer
on each magnet. To leave room for the gridded mechanical support structure,
depicted in  Fig.~\ref{fig:cad_structure}, adjacent drawers are separated by 
0.97~cm in the vertical dimension
and a minimum of 1.59~cm in the toroidal dimension.
Further details on the design of the support structure, as well as the 
finite-element calculations performed to qualify the structure for 
withstanding gravitational and internal magnetic forces, will be 
given in a separate paper \cite{zhu2022a}.

Design of the magnet layout was facilitated by the \textsc{Magpie} code
\cite{hammond2020a}, which was upgraded to accommodate the present concept
with cubic magnets.
The code can position magnets around any smooth toroidal surface, so the
layout scheme can be easily generalized to any stellarator with a 
plasma-conforming containment vessel. Parameters such as the thickness of
the magnet layer, the width of the gap spacing between adjacent magnets, and 
the number of toroidal wedges may be chosen by the user to best suit the 
physics requirements and engineering constraints. Magnets may also be excluded
from regions that would collide with other objects, such as toroidal-field
coils or access ports.

\section{Continuous magnet optimization}
\label{sec:optimization}

Once the positions of the magnets in the arrangement are specified, their
dipole moments are optimized to create the magnetic field shaping required
to confine a plasma in a given target equilibrium. For simplicity of 
fabrication, the number of unique magnet types is restricted: all magnets
are cubes with identical dimensions and one of a small number of polarization
orientations. Hence, the space of possible solutions is discrete and has a
large number of degrees of freedom that scales with the number of magnets
(of which there are typically tens of thousands). Finding an optimal solution 
within this large, discrete space is a challenging computational task.
Our approach to this problem divides the optimization procedure into a few 
steps. The first steps are continuous optimizations that take advantage of 
gradient-based algorithms. The continuous solutions are then projected into
the discrete space of allowable magnet types.  
This section describes the first steps
of the process, which employ continuous optimization.

The continuous optimizations were carried out by the \textsc{Famus} code
\cite{zhu2020b}. \textsc{Famus} models each magnet as an ideal dipole with
moment $\mathbf{m} = \rho M V \mathbf{\hat{v}}(\phi,\theta)$, where $\rho$ is a 
scaling constant, $M$ is the magnetization, $V$ is
the volume of the permanent magnet represented by the dipole, 
and $\mathbf{\hat{v}}(\phi,\theta)$ is a unit vector whose orientation is
specified by an azimuthal angle $\phi$ and polar angle $\theta$. The 
magnetization $M$ can be related to the remanent field $B_r$ of the permanent
magnet through $B_r~=~{\mu}M$, where $\mu$ is the permeability. To simplify
the optimization, any deviation of the magnets' permeability from that of
free space is neglected; hence, $\mu=\mu_0$. The
calculations in this paper assume a remanent field $B_r$~=~1.38~T, consistent 
with typical rare-Earth permanent magnets. Additional contributions to the
magnetic field from the toroidal field coils and from currents within the
target plasma were held fixed during the optimizations.

The continuous optimizations proceeded in two steps, distinguished primarily
by the objective function that the optimizer sought to minimize. In the 
first step, the parameters $\rho$, $\theta$, and $\phi$ for each dipole were
adjusted to minimize the magnetic field error, quantified by the surface 
integral

\begin{equation}
    \chi_B^2 = \iint_\mathcal{S}
        \left(\mathbf{B}\cdot\mathbf{\hat{n}}\right)^2dA, 
    \label{eqn:chi2b}
\end{equation}

\noindent where $\mathcal{S}$ is boundary of the target plasma, 
$\mathbf{\hat{n}}$ is the unit vector normal to $\mathcal{S}$, and
$\mathbf{B}$ is the total magnetic field evaluated on $\mathcal{S}$, including
contributions from the toroidal field coils, plasma current, and the
permanent magnets. Note that targeted magnetic field is perfectly achieved if
$\chi_B^2~=~0$. For each magnet, the scaling parameter $\rho$ was constrained
to remain between $-1$ and $1$, inclusive, such that the magnitude 
$|\mathbf{m}|$ of each dipole moment never exceeded $MV$. The parameter
$\rho$ was initialized to zero for each magnet, whereas the angles
$\theta$ and $\phi$ were initialized so that each dipole was approximately
aligned with the normal vector of a nearby point on the plasma vessel.

At the end of this first step, the solution should attain a certain minimum
level of field accuracy to be considered viable. For the studies in this
paper, we have chosen the following criterion for sufficient field accuracy:

\begin{equation}
    b_n = \frac{1}{B_0A}
              \iint_\mathcal{S}\left|\mathbf{B}\cdot\mathbf{\hat{n}}\right|dA
        \leq 0.005
    \label{eqn:bn}
\end{equation}

\noindent Here, $b_n$ is the normalized surface-averaged normal component of the
total magnetic field on the target plasma boundary, $A$ is the surface area
of the plasma boundary, and $B_0$ is the approximate
value of the magnetic field on the target plasma's magnetic axis, which is
0.5~T for our target plasma equilibrium. The choice of 
$b_n\leq0.005$ as the criterion was informed by free-boundary equilibrium
modeling of magnet solutions with a range of $b_n$ values. 
Should the optimized magnet solution fail to meet the criterion at this stage,
the geometric arrangement would need to be modified, typically by increasing
the thickness of the magnet layer and/or increasing the concentration of 
magnets close to the plasma.

If the first step succeeds in meeting the criterion, a
second step is performed to find a refined solution in which the optimized
values of $\rho$ have an absolute value of either 1 or 0.
Attaining such a binary distribution of $|\rho|$
for the set of dipoles is an important step on the way to a discrete
solution. If every dipole has either zero strength or full strength 
$|\mathbf{m}|=MV$, the solution can be realized by placing magnets with
a uniform volume $V$ at the locations of dipoles with $|\rho|$=1 and by 
simply leaving empty the locations of dipoles with $\rho=0$. If, on the
other hand, the solution had many dipoles with intermediate
$|\rho|$ values, it would be necessary to use magnets with 
a wide range of volumes $V$, which is incompatible with the requirement of using
magnets with identical dimensions.

To this end, the dipole moment parameters are put through a second continuous
optimization that minimizes a linear
combination of two objective functions: $\chi_B^2 + \lambda\chi_\rho^2$, where

\begin{equation}
    \label{eqn:chi2rho}
    \chi_\rho^2 = \sum_i^N \left(|\rho_i|(1 - |\rho_i|)\right)^2,
\end{equation}

\noindent $N$ is the number magnets whose dipole moments are to be optimized,
$i$ is the index number for each magnet, and $\rho_i$ is the scaling constant
for the moment magnitude of the $i^{th}$ magnet. As can be seen in 
Eq.~\ref{eqn:chi2rho}, the objective function $\chi_\rho^2$ favors magnets
with $\rho=0$ or $|\rho|=1$ and penalizes magnets with intermediate values
of $|\rho|$. The weighting factor $\lambda$, which is held fixed during the
optimization, is chosen so that the solution
retains good field accuracy while eliminating the presence of magnets with
intermediate moment magnitudes. In practice, the optimization must be repeated
with different values of $\lambda$ until a suitable trade-off is found.

\begin{figure*}
    \begin{center}
    \includegraphics[width=\textwidth]{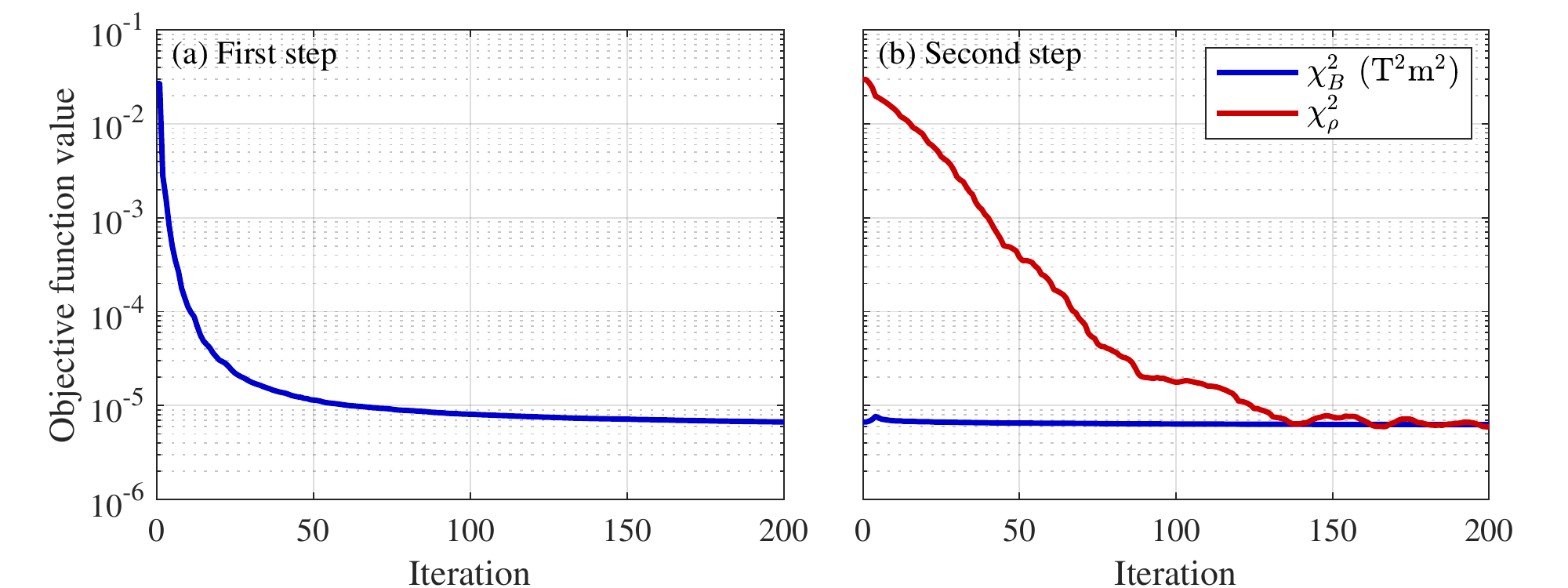}
    \caption{Values of the objective functions $\chi_B^2$ and $\chi_\rho^2$
             during a two-step continuous optimization of the dipole moments
             of the magnet arrangement shown in Fig.~\ref{fig:mag_argmt}
             for the target plasma.
             (a) first step, in which only $\chi_B^2$ was minimized;
             (b) second step, which continued from the end point of the
                 first step while minimizing a linear combination of 
                 $\chi_B^2$ and $\chi_\rho^2$.}
    \label{fig:opt_plots}
    \end{center}
\end{figure*}

Traces of the objective functions from an optimization of the dipole moments
of the magnets in the arrangement from Fig.~\ref{fig:mag_argmt} are shown in
Fig.~\ref{fig:opt_plots}. The dipole moments in this case were optimized for
the target plasma, and the objective function $\chi_B^2$ was 
computed accordingly (Eq.~\ref{eqn:chi2b}). In the first step, in which
the optimizer seeks to exclusively minimize field error, $\chi_B^2$ decreased
by almost four orders of magnitude (Fig.~\ref{fig:opt_plots}a). In the 
second step, in which a weighted sum of $\chi_B^2$ and $\chi_\rho^2$ is 
minimized, $\chi_\rho^2$ undergoes a reduction of nearly four orders of 
magnitude while avoiding an increase of $\chi_B^2$, which in fact improves 
slightly (Fig.~\ref{fig:opt_plots}b).

\begin{figure}
    \begin{center}
    \includegraphics[width=0.5\textwidth]{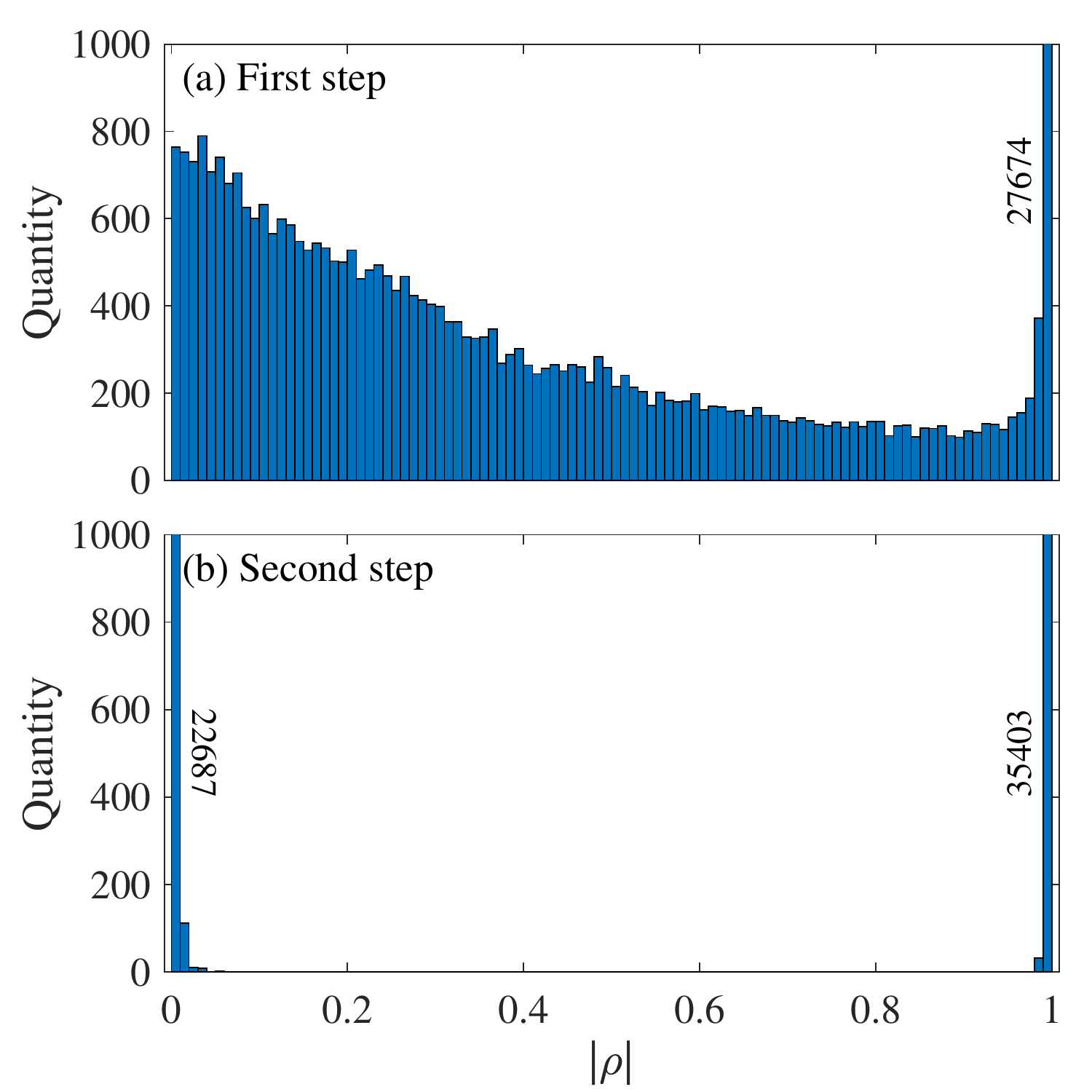}
    \caption{Histograms of normalized dipole moment magnitude $|\rho|$ for
             a set of magnets following (a) the first step and (b) the second
             step of a continuous optimization of the magnets in the arrangement
             shown in Fig.~\ref{fig:mag_argmt} for the target plasma.
             Bin quantities that exceed the y-axis limit are written alongside
             the respective bin.}
    \label{fig:rho_distro}
    \end{center}
\end{figure}

The effects of the two optimization steps on the distribution of dipole
moment magnitudes can be seen in Fig.~\ref{fig:rho_distro}. After the first
step (Fig.~\ref{fig:rho_distro}a), just under half of the 58,258 dipoles in the 
arrangement have attained the maximum moment magnitude allowed by the 
optimization procedure, corresponding to $|\rho|=1$. The rest of the 
distribution is broadly spread among intermediate values of $|\rho|$. Then, 
after the second step (Fig.~\ref{fig:rho_distro}b), the intermediate values of 
$|\rho|$ have largely been removed from the distribution, with all dipole 
moments having a magnitude of nearly zero or the maximum value. 

\section{Mapping dipole moments to a discrete subset}
\label{sec:mag_mapping}

Following the continuous optimization steps in Sec.~\ref{sec:optimization},
the dipole moments in the magnet array exhibit a nearly binary distribution of
magnitude, but the directions of each vector are unrestricted. To limit the
number of polarization types required in the array, the directions must be
adjusted to the nearest direction permitted by a discrete set of polarization
types. Since this adjustment entails moving the vectors away from their 
optimal orientations, it will reduce the magnetic field accuracy of the 
array. Hence, using a larger number of polarization types would require 
smaller adjustments to the optimized moment directions and therefore enable
greater field accuracy. However, the need for field accuracy must be balanced 
against the added complexity of fabricating magnets with many different 
polarization types.

This section describes the assessment and selection of polarization types for
the discrete solution. Sec.~\ref{sec:types} describes the six types considered
in this work. Sec.~\ref{sec:pol_combos} describes the different subsets of 
those six types that were evaluated for discrete solutions. 
Sec.~\ref{sec:disc_pm4stell} presents a comparison of discrete solutions
for the target plasma employing the different subsets of types, with a
focus on the subset of three types (which will be identified as ``Subset 5'' in 
Sec.~\ref{sec:pol_combos}) that we ultimately chose for the 
magnet array design.

\subsection{Magnet polarization types}
\label{sec:types}

\begin{figure}
    \begin{center}
    \includegraphics[width=0.48\textwidth]{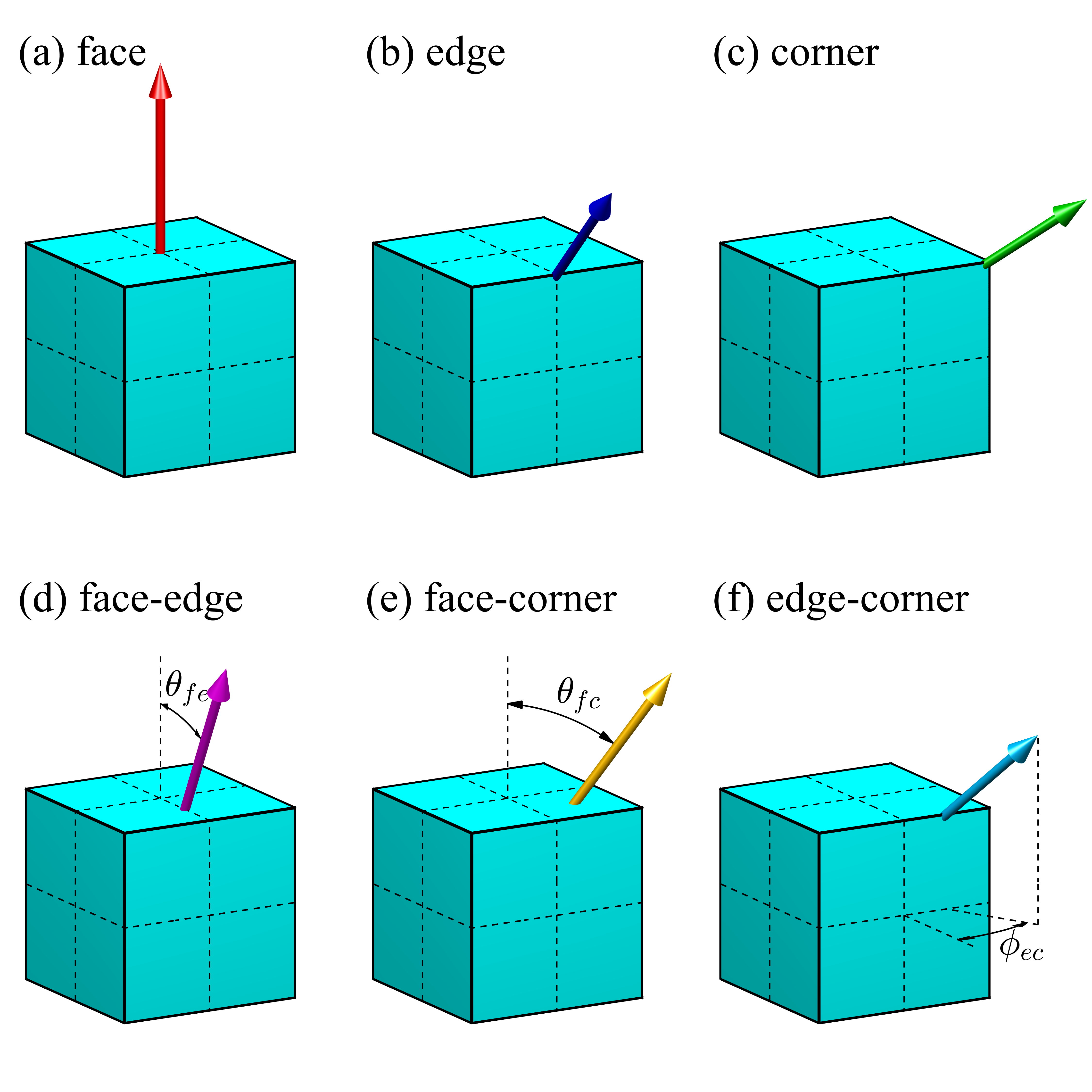}
    \caption{Polarization orientations relative to the cubic geometry of the
             magnet types considered for this study.}
    \label{fig:pol_types}
    \end{center}
\end{figure}

The six polarization types considered in this work are shown in
Fig.~\ref{fig:pol_types}. The \textit{face} type (Fig.~\ref{fig:pol_types}a)
exhibits a magnetization represented by a unit vector \uvec[f]{v} 
perpendicular to the planes of two opposite faces. 
This type admits six distinct dipole moment vectors from the possible 
rotations of the cube that preserve the cube's geometric orientation.
The \textit{edge} type (Fig.~\ref{fig:pol_types}b) 
exhibits a magnetization vector \uvec[e]{v} such that a line traced from
the cube's centroid in this direction would intersect the midpoint of an edge.
This type admits twelve different polarizations. The \textit{corner} type 
(Fig.~\ref{fig:pol_types}c) has a polarization vector \uvec[c]{v} such that
a line traced from the centroid in this direction would intersect one of the
cube's corners. It admits eight different polarizations.

We also considered three ``hybrid'' polarization types that lie in planes
defined by  
pairs of the basic types introduced above. The \textit{face-edge} type
(Fig.~\ref{fig:pol_types}d) has a polarization vector \uvec[fe]{v} that is 
coplanar with a face-type unit vector \uvec[f]{v} and an edge-type unit vector
\uvec[e]{v}, and is specified by the angle 

\begin{equation}
    \label{eqn:theta_fe}
    \theta_\text{fe} = \arccos(\uvec[fe]{v} \cdot \uvec[f]{v})
\end{equation}

\noindent to the nearest face-type vector. Assuming 
$0<\theta_\text{fe}<45^\circ$, there are twenty-four distinct polarization 
vectors obtainable through different rotations of the cube. The 
\textit{face-corner} type (Fig.~\ref{fig:pol_types}e) has a 
polarization vector \uvec[fc]{v} that is coplanar with a face-type unit 
vector \uvec[f]{v} and a corner-type unit vector 
\uvec[c]{v}, and is specified by the angle 

\begin{equation}
    \label{eqn:theta_fc}
    \theta_\text{fc} = \arccos(\uvec[fc]{v} \cdot \uvec[f]{v})
\end{equation}

\noindent to the nearest face-type vector. This admits twenty-four 
distinct vectors through rotations if 
$0<\theta_\text{fc}<\arccos\left(1/\sqrt{3}\right)\approx54.7^\circ$.
Finally, the \textit{edge-corner} type (Fig.~\ref{fig:pol_types}f) lies in a 
plane parallel to an edge-type vector \uvec[e]{v} and a corner-type vector
\uvec[c]{v}, and is specified by the azimuthal displacement angle 
$\phi_\text{ec}$ that may be expressed as

\begin{equation}
    \label{eqn:phi_ec}
    \phi_\text{ec} = \arcsin\left(
        \frac{\vect[e]{p}\times\vect[ec]{p}}{|\vect[e]{p}||\vect[ec]{p}|} 
        \right),
\end{equation}

\noindent where \vect[e]{p} and \vect[ec]{p} are the projections of \uvec[e]{v}
and \uvec[ec]{v} onto the plane perpendicular to the adjacent face-type
vector \uvec[f]{v} for which $\uvec[f]{v}\cdot(\uvec[e]{v}\times\uvec[ec]{v})$
is positive. Assuming $0<\phi_\text{ec}<45^\circ$, there are twenty-four
attainable polarization vectors \uvec[ec]{v} from different rotations of the
cube, as with \uvec[fe]{v} and \uvec[fc]{v}. 

\subsection{Subsets of polarization types}
\label{sec:pol_combos}

A major choice that must be made in this design process is which, and how many, 
of the magnet polarization types to admit in the discrete solution. In general,
employing more types implies more possible polarization vectors for each
position in the magnet array, admitting solutions that can attain higher 
field accuracy and utilize the magnet material more efficiently. On the other
hand, a larger number of polarization types leads to greater design complexity.
Furthermore, certain polarization types are substantially more expensive to 
fabricate than others, depending primarily on how many of the cube's faces the 
polarization aligns with. For example, the face-type polarization is the 
simplest and cheapest to fabricate, whereas the edge-corner type would be the 
most complex and expensive. Hence, the choice of types to utilize needs to
balance the physics requirement for field accuracy against the costs of
fabrication.

\begin{table*}
    \begin{center}
    \begin{tabular}{|c|c|c|c|c|c|c|c|c|c|c|c|c|}
        \hline
        Subset & F & E & C & FE & FC & EC & $\theta_\text{fe}$
            & $\theta_\text{fc}$ & $\phi_\text{ec}$ 
            & \makecell{Num.~of\\polarizations}
            & \makecell{Max.~$\theta_\text{offs}$ \\ (deg.)}
            & \makecell{$\rmsoffs$\\ (deg.)} \\
        \hline
        1 & \checkmark & & & & & & & & & 6 & 54.7 & 34.0 \\
        2 & \checkmark & \checkmark & & & & & & & & 18 & 35.2 & 19.9 \\
        3 & \checkmark & \checkmark & \checkmark & & & & & & & 26 & 27.5 
          & 16.6 \\
        4 & \checkmark & \checkmark & & \checkmark & & & 22.5 & & & 42 & 35.2 
          & 16.2 \\
        5 & \checkmark & & & \checkmark & \checkmark & & 30.3 & 38.1 & & 54 
          & 19.4 & 11.3 \\
        6 & & \checkmark & & \checkmark & \checkmark & & 18.4 & 38.6 & & 60
          & 18.4 & 10.6 \\
        7 & \checkmark & \checkmark & & \checkmark & \checkmark & & 22.5 & 38.6
          & & 66 & 16.9 & 10.2 \\
        8 & \checkmark & \checkmark & \checkmark & \checkmark & \checkmark
          & \checkmark & 22.5 & 27.0 & 22.5 & 98 & 14.1 & 8.50 \\
        \hline
    \end{tabular}
    \caption{Properties of the subsets of polarization types considered in
             this work. Check marks indicate the presence of a given type
             in each subset. The types are abbreviated as follows: F=face,
             E=edge, C=corner, FE=face-edge, FC=face-corner, and EC=edge-corner.
             Angular parameters for the hybrid types ($\theta_\text{fe}$, 
             $\theta_\text{fc}$, $\phi_\text{ec}$), as defined in 
             Sec.~\ref{sec:types} and Fig.~\ref{fig:pol_types}, are given if 
             their respective types are included in the subset.}
    \label{tab:pol_sets}
    \end{center}
\end{table*}

To inform this choice, eight different subsets of the polarization types were
evaluated for their attainable field accuracy. The subsets are summarized
in Table~\ref{tab:pol_sets}. The subsets ranged from the simplest possible,
containing only the face-type polarization (Set 1) to containing all six
types (Set 8). 

The number of polarizations in Table~\ref{tab:pol_sets} indicates the number 
of unique dipole moment vectors that can be created at a given position in the 
magnet array, given a choice of magnets of the types of polarizations within 
the subset and considering all possible orientations of the cube that maintain 
its geometric footprint. For example, when using Subset 1, which only contains
magnets with the face-type polarization, six different dipole moment vectors 
can be created at a given position in the array. When using Subset 2, by 
contrast, either a face-type or an edge-type polarized magnet may be placed
at each position. This yields a total of eighteen possible diople moment
vectors (six for the face type and twelve for the edge type).

The discretization procedure described here entails rotating 
continuously-optimized dipole moment vectors to the nearest possible vectors
from a discrete subset. Thus, the dipole moments output from the discretization
procedure will have an angular offset,
$\theta_\text{offs}$, from the original direction. Correspondingly, a given
subset of polarization types can be characterized by its distribution of
$\theta_\text{offs}(\phi,\theta)$ over the unit sphere, parametrized by
the azimuthal angle $\phi$ and polar angle $\theta$. Formally, 
$\theta_\text{offs}$ is the distribution of the angular separation from the 
nearest allowable dipole moment from the set of polarization types:

\begin{equation}
    \theta_\text{offs}(\phi,\theta) = \min_{\uvec[s]{v}\in{S}} 
        \left[ \arccos\left(\uvec{v}(\phi,\theta) \cdot \uvec[s]{v} \right) 
            \right]
    \label{eqn:theta_offs}
\end{equation}

\noindent Here, $S$ contains all available polarization unit vectors 
\uvec[s]{v} within a given subset of polarization types. 
$\uvec{v}(\phi,\theta)$ is a unit vector representing the direction of an 
arbitrary, continuously-optimized dipole moment.

\begin{figure*}
    \begin{center}
    \includegraphics[width=\textwidth]{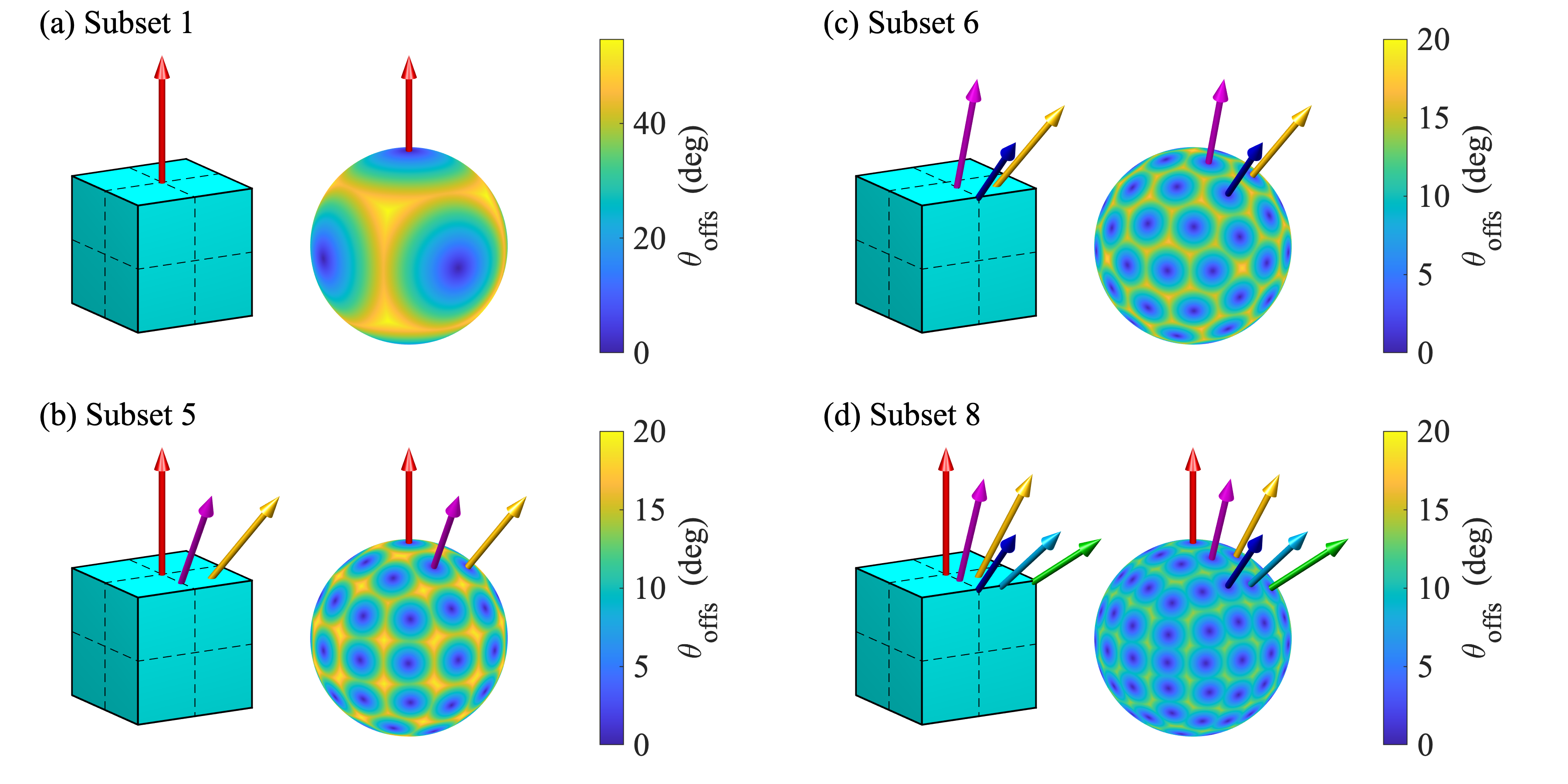}
    \caption{Example polarization vectors relative to a cubic magnet (left
             side of each subfigure) and distribution of $\theta_{\text{offs}}$ 
             over the unit
             sphere (right side) for a selection of the polarization type 
             subsets specified in Table~\ref{tab:pol_sets}.
             Note that the color scale for Subset 1 (a) is different from that
             of the others (b)-(d).}
    \label{fig:pol_sets}
    \end{center}
\end{figure*}

Distributions of $\theta_\text{offs}$ from a few subsets of polarization types
are shown in Fig.~\ref{fig:pol_sets}. The distributions are each plotted on
spherical domains. For visual reference, an example of each type of polarization
vector in the set is also shown. In addition, the reference vectors are 
shown relative to the geometry of a cubic magnet to the left of each sphere.
Each allowable polarization vector within a subset appears as a local minimum
of $0^\circ$ in the distribution of $\theta_\text{offs}$. Naturally, the
greatest values of $\theta_\text{offs}$ are seen for the subsets with the 
fewest allowable polarization vectors.

The distribution of $\theta_\text{offs}$ can be used to define two figures
of merit for each subset of polarization types. The first is simply the maximum
value of $\theta_\text{offs}$ found on the unit sphere for a given subset.
The second is the root-mean-square of $\theta_\text{offs}$ over the domain:

\begin{equation}
    \left\lVert \theta_\text{offs} \right\rVert_2 = 
        \sqrt{\frac{1}{4\pi} \int_0^{\pi} \int_0^{2\pi} 
            \theta_\text{offs}^2~\sin\theta~d\phi~d\theta}
    \label{eqn:ms_theta_offs}
\end{equation}

\noindent Both of these quantities are shown for each set in 
Table~\ref{tab:pol_sets}. They were estimated numerically on a 
grid of $\phi$ and $\theta$ values; the former by simply finding the maximum
value of $\theta_\text{offs}$ on the grid, and the latter through numerical
integration. 

Note that these quantities are functions of the polarization type subsets
alone and are independent of the properties of any magnet array or target
plasma. Nevertheless, it is reasonable to expect that subsets with lower values
of \rmsoffs\ will generally introduce less error during the discretization 
procedure. Hence, \rmsoffs\ helped to inform the design of some of
the subsets. In particular, $\theta_\text{fe}$ and $\theta_\text{fc}$ for 
Subsets 5 and 6 were both chosen to minimize $\rmsoffs$.



\subsection{Discretized solutions}
\label{sec:disc_pm4stell}

\begin{figure}
    \begin{center}
    \includegraphics[width=0.48\textwidth]{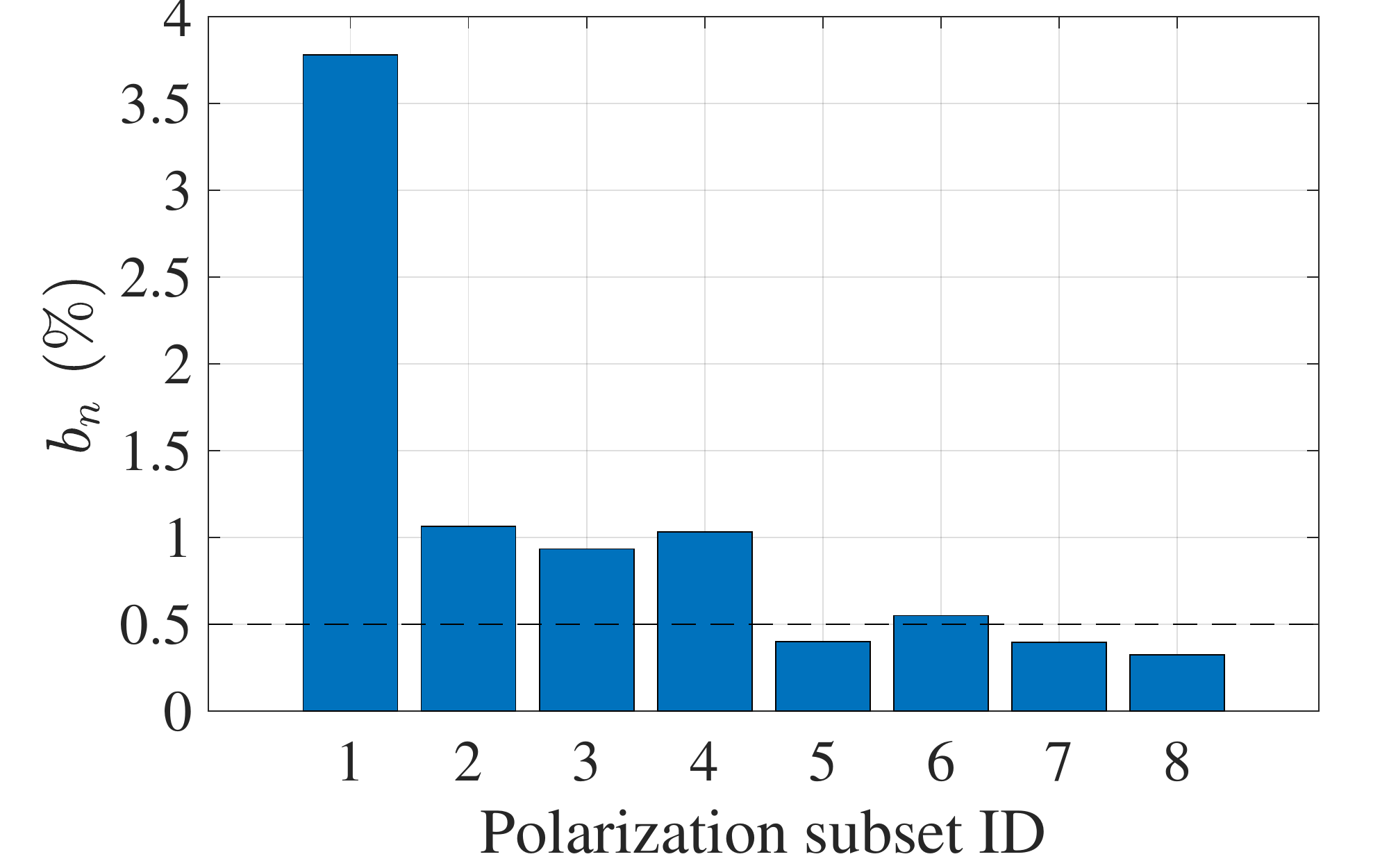}
    \caption{Value of $b_n$ (Eq.~\ref{eqn:bn}) on the boundary of the target
             plasma equilibrium after rotating the continuously-optimized 
             dipole moments of the magnet array to each of the sets of 
             polarization types specified in Table~\ref{tab:pol_sets}.
             The horizontal dashed line indicates the criterion for field
             accuracy defined in Eq.~\ref{eqn:bn}.}
    \label{fig:bn_by_set}
    \end{center}
\end{figure}

The continuously-optimized dipole moments for the target plasma were 
discretized to polarizations from each of the eight subsets summarized in
Table~\ref{tab:pol_sets}. The magnetic field error of each of these discretized
solutions, indicated by $b_n$ (Eq.~\ref{eqn:bn}), is shown in 
Fig.~\ref{fig:bn_by_set}. 
The predominant trend is that $b_n$ tends to decrease as the number of 
available polarizations increases, and as the \rmsoffs\ metric decreases. This 
is to be expected, as sets with more polarization options and lower offset 
angles tend to enable discrete solutions that are closer to the 
continuously-optimized solutions that they approximate. On the other hand, the 
dependence of $b_n$ on \rmsoffs\ is not perfectly monotonic: for example, 
Subset 6 exhibits a lower \rmsoffs\ than Subset 5 (Table~\ref{tab:pol_sets}), 
whereas the discretized solution utilizing Subset 5 attains a lower value of 
$b_n$. Note that overall, Subsets 5, 7, and 8 have satisfied the criterion
for field accuracy defined in Eq.~\ref{eqn:bn}.

\begin{figure*}
    \begin{center}
    \includegraphics[width=\textwidth]{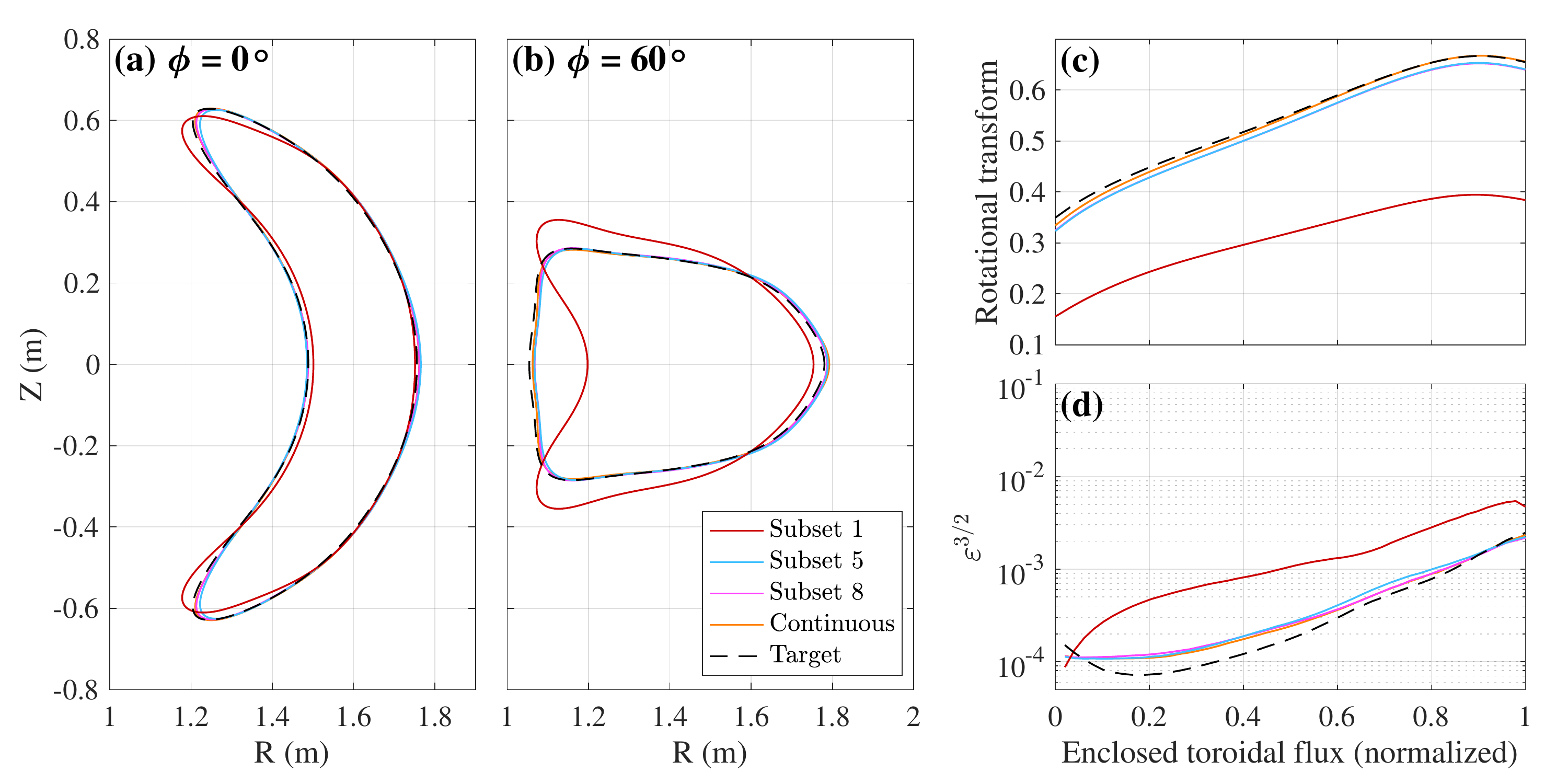}
    \caption{Equilibrium properties of plasmas confined with the fields
             generated by selected magnet solutions as compared with the 
             targeted plasma equilibrium. 
             (a) Cross-sections of the plasma boundary at toroidal angle
                 $\phi=0^\circ$;
             (b) Cross-sections of the plasma boundary at toroidal angle
                 $\phi=60^\circ$;
             (c) Profiles of the rotational transform; and
             (d) Profiles of the effective ripple.}
    \label{fig:equil_by_set}
    \end{center}
\end{figure*}

\begin{figure}
    \begin{center}
    \includegraphics[width=0.48\textwidth]{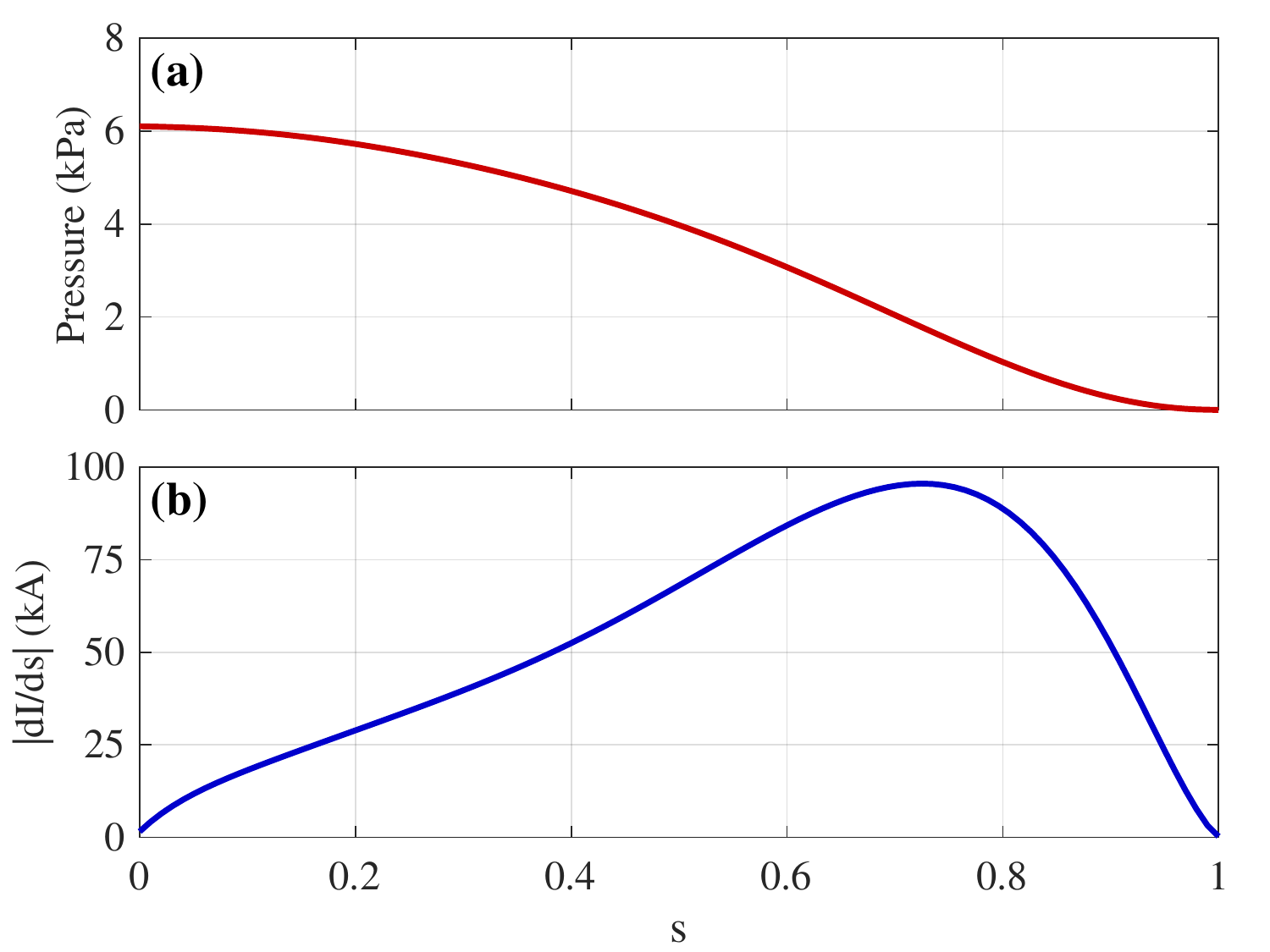}
    \caption{Profiles of (a) plasma pressure and (b) the radial derivative of 
             toroidal plasma current as a function of normalized toroidal 
             flux $s$ for the target plasma equilibrium.}
    \label{fig:pi}
    \end{center}
\end{figure}

For a more detailed look at the suitability of the different solutions for
confining the target plasma, we performed free-boundary magnetohydrodynamic
(MHD) equilibrium calculations with the VMEC code 
\cite{hirshman1983a,hirshman1986a}. Each calculation assumed plasmas
with the same prescribed toroidal current profile, pressure profile, and
enclosed toroidal flux as the target plasma. The profiles of pressure and
current are shown Fig.~\ref{fig:pi}. The current profile is based on a 
neoclassical estimate of the bootstrap current for the target plasma
\cite{zarnstorff2001a}. The external magnetic fields
for each plasma included contributions from the NCSX toroidal field coils
and the permanent magnets, with each permanent magnet approximated 
as an ideal dipole. All modeled equilibria had volume-averaged $\beta$ values of
about $4.1\%$, similar to the target plasma. After the equilibrium properties 
were determined,
the profile of effective ripple, a measure of neoclassical transport, was
calculated using the NEO code \cite{nemov1999a}. 

Results of these calculations are shown in Fig.~\ref{fig:equil_by_set} for
discrete solutions utilizing Subsets 1, 5, and 8, as well as the solution with
continuous polarization angles used as input for the discretization.
Key equilibrium properties are compared with those of the targeted 
plasma, including boundary geometry (Fig.~\ref{fig:equil_by_set}a-b),
rotational transform profile (Fig.~\ref{fig:equil_by_set}c), and
effective ripple profile (Fig.~\ref{fig:equil_by_set}d). As expected, the
worst agreement is found for the discrete solution using Subset 1,
which contains only face-type magnets and therefore exhibits the
greatest offsets from the continuously optimized dipole moments. By contrast,
the discrete solutions with Subsets 5 and 8 yield plasma equilibria that are
quite similar with that of the continuous solution and exhibit characteristics
much closer to those of the targeted plasma. In particular, both Subsets 5 and
8 attain rotational transform profiles that remain within 0.03 of the target
values, and their effective ripple, indicated by the metric 
$\epsilon_\text{eff}^{3/2}$, remains close to $10^{-4}$ at the core and 
$10^{-3}$ at the edge.

As indicated by the free-boundary equilibrium calculations in 
Fig.~\ref{fig:equil_by_set}, Subset~8 does not yield a major improvement in 
field accuracy over Subset~5 despite having lower values of $\rmsoffs$.
On the other hand, Subset~5 would be substantially simpler to fabricate
by virtue of consisting of half as many unique polarization types. For this 
reason, we adopted Subset~5 for further design, fabrication, and construction 
activities.

\begin{figure*}
    \begin{center}
    \includegraphics[width=\textwidth]{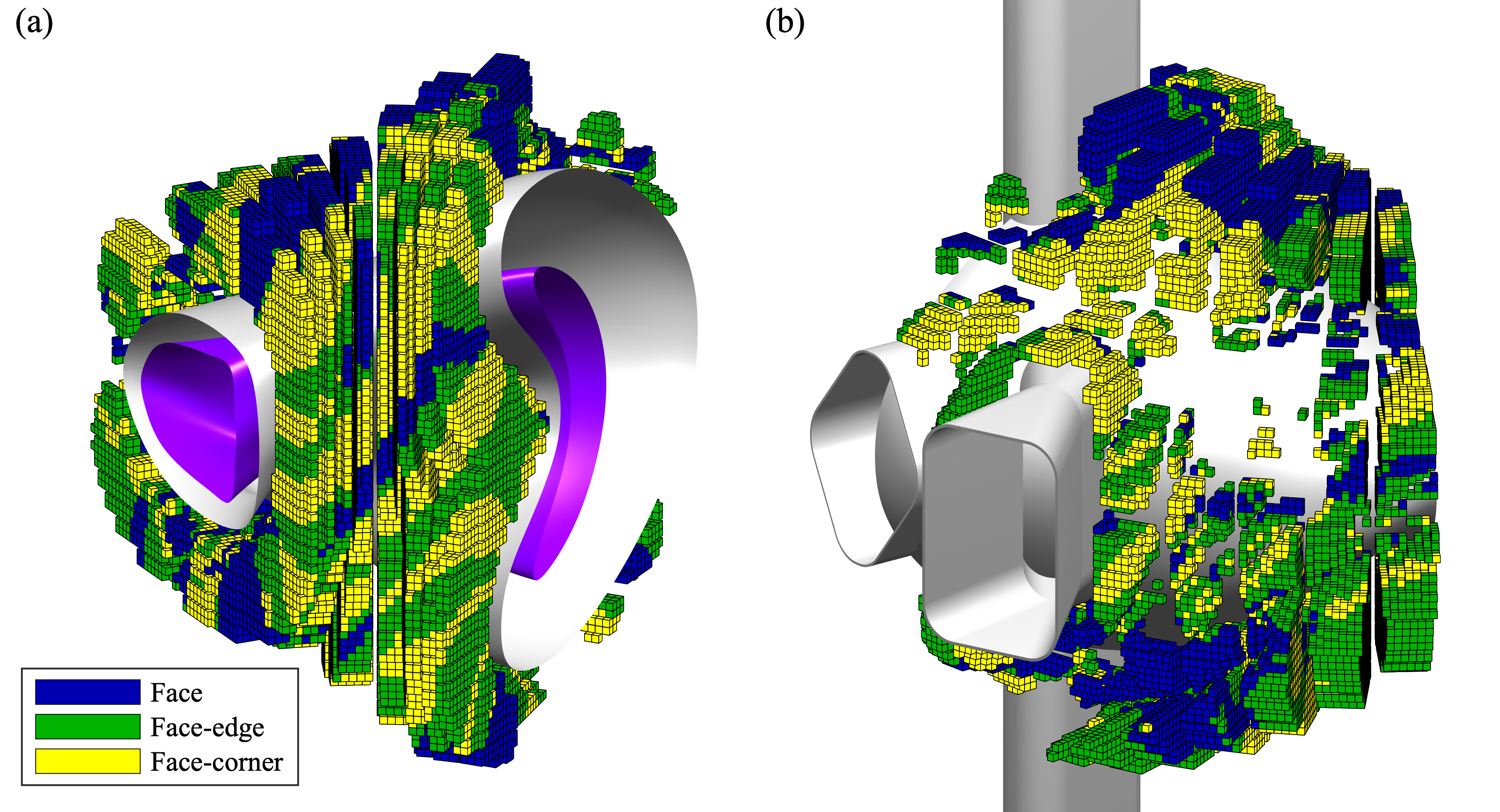}
    \caption{Renderings of a discretized magnet solution for the target
             plasma, viewed from (a) the inboard side and (b) the outboard
             side. The cubic magnets are color-coded according to polarization
             type. Magnets with optimized dipole moments of magnitude zero, 
             most of which were on the outboard side,
             have been omitted from the initial arrangement. The boundary of 
             the plasma is also shown in purple.}
    \label{fig:magnets_pm4stell}
    \end{center}
\end{figure*}

The discretized solution using Subset 5---consisting of the face, face-edge, and
face-corner polarization types---is shown in Fig.~\ref{fig:magnets_pm4stell}.
The geometric layout of the magnets corresponds to the initial arrangement 
shown in 
Fig.~\ref{fig:mag_argmt}, omitting the magnets that were given dipole moments
of zero during the optimization procedure. Overall, the solution for one
half-period contains 35,436 cubic magnets. These include 8,778 of the face 
polarization type (24.8\%), 13,735 of the face-edge type (38.8\%), and
12,923 of the face-corner type (36.5\%). Extending this solution around the
full stellarator (six half-periods), therefore, would require 212,616 magnets, 
constituting a magnet volume of 5.74~m$^3$.

\begin{figure}
    \begin{center}
    \includegraphics[width=0.48\textwidth]{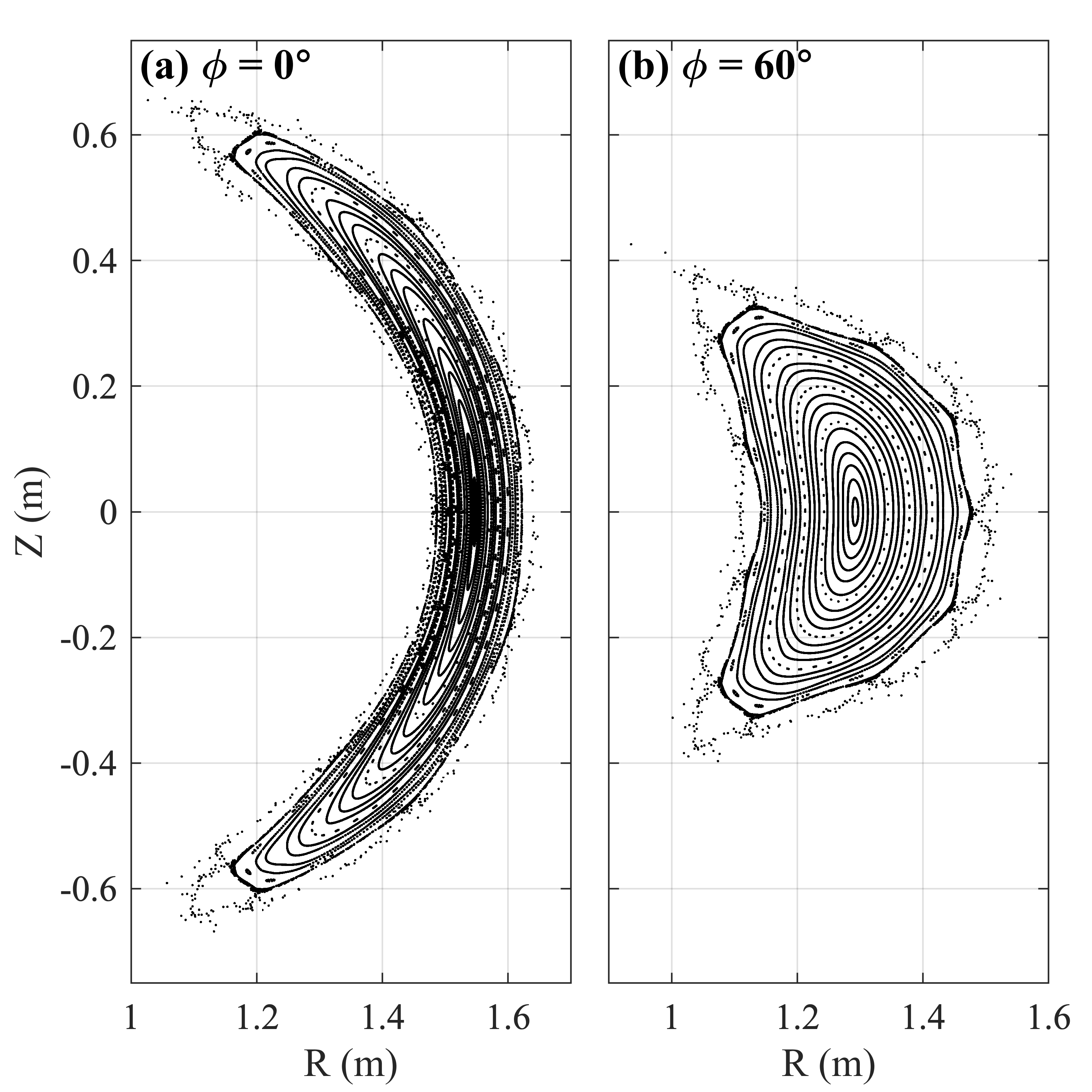}
    \caption{Cross-sections of magnetic field lines traced in the vacuum field
             produced by the magnets from the discretized solution with 
             polarizations from Subset~5 and the NCSX TF coils at 
             toroidal angles (a) $\phi=0^\circ$ and (b) $\phi=60^\circ$.}
    \label{fig:poincare}
    \end{center}
\end{figure}

An important practical consideration for stellarator design is whether the
magnets can produce closed flux surfaces in the absence of plasma or plasma
current. Vacuum flux surfaces greatly ease the requirements for plasma 
startup and operation at low beta, which are necessary conditions for reaching
the target plasma equilibrium. Fortunately, the vacuum field from Subset~5 and 
the NCSX TF coils does indeed contain flux surfaces. This can be seen in 
Fig.~\ref{fig:poincare}, which shows Poincar\'{e} cross-sections of magnetic
field lines at two toroidal positions. The \textsc{Fieldlines} code 
\cite{lazerson2016a} was employed to compute the field line trajectories.

\section{Conclusions and future work}
\label{sec:conclusions}

In this work, we have produced a design for an array of cubic magnets of 
three distinct types that confines a quasi-axisymmetric stellarator plasma.
Our approach divided the task into three principal steps: (1) specification of
the allowable positions, shapes, and geometric orientations of the magnets;
(2) continuous optimization of the magnets' dipole moments; and (3) adjustment
of the optimized dipole moments to conform to certain allowable polarization
types. Several subsets of polarization types were considered, and ultimately a
subset consisting of three distinct types was chosen that exhibited high field
accuracy while limiting the complexity of fabrication.

The design of a magnet array for a variant of an NCSX plasma equilibrium was 
the primary motivating factor for the work described in this paper. 
However, the procedures developed here are broadly generalizable to any 
stellarator plasma configuration, as long as the magnitudes of the required 
shaping fields are within the capabilities of existing permanent magnet 
materials. Whether or not rare-Earth magnets are capable of confining a
given target plasma depends on its magnetic and structural requirements.
The magnetic requirements include the on-axis
toroidal field strength, the amount of external poloidal field shaping 
required for a given on-axis field strength, and whether or not toroidal-field
coils contribute to the external poloidal field (as is the case with, for 
example, the tilted coils used in the concept proposed in 
Ref.~\cite{helander2020a}). Key structural considerations include the amount 
of spacing required between the plasma vessel and the magnet array and the 
amount of spacing required between the magnets themselves for structural 
supports.

The optimization procedure developed in this work achieves a similar end goal
to that developed by Lu et al.~\cite{lu2022a} for designing stellarators with
cubic magnets with discrete polarizations. A common aspect of both approaches 
is that they entail optimizing the dipole moments of an arrangement of magnets 
with prescribed locations and spatial orientations. However, the optimization
procedures differ. In Ref.~\cite{lu2022a}, the dipole moments are
optimized in two stages: first, the dipole moment of each magnet is iteratively
adjusted to minimize the normal field at a nearby location on the plasma
boundary; and second, the iterative adjustments on each magnet are repeated
to minimize the global residual quantity $\chi_B^2$ (as defined in 
Eq.~\ref{eqn:chi2b}). The iterative adjustments for each magnet involve 
cycling through the allowable discrete polarization vectors and choosing the
one that yields the best result; hence, the approach is a fully discrete 
optimization procedure. By contrast, the approach described in this paper
entails initial steps that utilize continuous optimization procedures 
(Sec.~\ref{sec:optimization}) to yield a solution with a continuous distribuiton
of dipole moment directions, which are then adjusted to the nearest of a
set of discrete vectors defined for each magnet (Sec.~\ref{sec:mag_mapping}).
In addition, the procedure developed here only considers the global residual
field error quantity $\chi_B^2$ and does not include a stage that optimizes 
each magnet according to its local contribution to the field at the plasma 
boundary.

In this design study, the dipole moment optimizations and calculations of field 
accuracy treated each magnet as an ideal magnetic dipole, contributing in a
fixed and linear way to the overall magnetic field distribution. 
In practice, the actual field generated by the magnet set will differ from the 
idealized calculations, both as a result of the permeability of the magnets 
and, to some extent, slight demagnetization of some of the magnets in response 
to the external magnetic field. Furthermore, if such an array is to be
fabricated and constructed, additional field errors will arise from 
imperfections in fabrication and spatial misalignments of the magnets within
the mounting structure.

To correct for misalignments and nonlinear field effects, a secondary array 
of error-correction magnets is foreseen to be placed between the main magnet
array and the plasma vessel. The main magnet array described in this paper
was constrained to allow room for this additional set of magnets. This 
error-correcting array will be specified following the construction and 
assembly of the main magnet array, at which point the deviations of the field
of the as-built array from the ideal prediction can be measured. The 
required dipole moments for these magnets can then be determined using the
same procedure outlined in this paper for the main magnets. As will be
described in detail in forthcoming publications, assessments of the sensitivity
of the magnetic field to positional offsets \cite{chambliss2022a}, 
simulations of possible positional magnet offsets \cite{rutkowski2022a}, and 
finite-element models of the magnet array that account for finite magnet size,
permeability, and demagnetization \cite{zhu2022a}
have indicated that all of these deviations should be within the correction 
capabilities of the secondary array.

\section*{Acknowledgments}

This work was supported by the Advanced Research Projects Agency -- Energy
(ARPA-E), U.S. Department of Energy, under contract number DE-AC02-09CH11466. 
The views and opinions of authors expressed herein do not necessarily state or
reflect those of the U.S. Government or any agency thereof. The U.S. Government 
retains a non-exclusive, paid-up, irrevocable, world-wide license to publish or 
reproduce the published form of this manuscript, or allow others to do so, for 
U.S. Government purposes.

\section*{Conflict of interest}

A patent application (US 63/319,568) incorporating parts of this work has been
filed by Princeton University and authors K.C.H., C.Z., and D.A.G.

\section*{Data and code availability}

The data presented in this paper are available at 
\url{http://arks.princeton.edu/ark:/88435/dsp01x059cb547}.
The VMEC, NEO, and \textsc{Fieldlines} codes are part of the 
\textsc{Stellopt} suite of codes \cite{stellopt} and may be accessed at
\url{https://github.com/PrincetonUniversity/STELLOPT}.
Access to the \textsc{Magpie} and \textsc{Famus} codes may be given upon 
request.


%

\end{document}